\documentclass[%
 reprint,
 superscriptaddress,
 amsmath,amssymb,
 aps,
floatfix,
longbibliography]{revtex4-1}

\usepackage{graphicx}
\usepackage{bm}

\usepackage{xcolor}
\usepackage[colorlinks=true,citecolor=blue,linkcolor=magenta]{hyperref}

\usepackage{verbatim}

\newcommand{\nn}{\nonumber}

\newcommand{\rar}{\rightarrow}
\newcommand{\da}{\downarrow}
\newcommand{\ua}{\uparrow}
\newcommand{\ben}{\begin{eqnarray}}
\newcommand{\een}{\end{eqnarray}}

\newcommand{\be}{\begin{equation}}
\newcommand{\ee}{\end{equation}}

\begin{document}
\title{Wei-Norman approach for non-Hermitian driven spin-$S$ systems and its application to defect freezing}

\author{Mingwei Meng}
\affiliation{School of Physics and Electronics, Hunan University, Changsha 410082, China}
\author{Chen Sun}
\email{chensun@hnu.edu.cn}
\affiliation{School of Physics and Electronics, Hunan University, Changsha 410082, China}

\begin{abstract}

In the theoretical study of nonequilibrium non-Hermitian systems, obtaining exact analytical solutions for their nonadiabatic dynamics is highly desirable yet often challenging. In this work, we identify a class of non-Hermitian quantum systems where 
this difficulty can be substantially reduced. Employing the Wei-Norman approach, we show that for a spin-$S$ subject to a general non-Hermitian time-dependent drive, the matrix elements of the evolution operator can be expressed in closed analytical forms (via Jacobi polynomials) in terms of the corresponding spin-$1/2$ model. This approach is straightforward and accessible to nonspecialists in Lie algebra. As an application, we investigate a specific nonequilibrium non-Hermitian phenomenon known as defect freezing, i.e., the existence of excitations in the adiabatic limit, in spin-$S$ extensions of the $\mathcal{PT}$-symmetric Su-Schrieffer-Heeger model under linear quenches. We derive exact analytical expressions for the momentum-resolved excitation probabilities and the total excitation densities. Our results reveal that defect freezing occurs exclusively in momentum sectors that traverse the $\mathcal{PT}$-symmetry-broken region---and thus pass through a pair of higher-order exceptional points (EPs)---during the quench; notably, the excitation density exhibits a singularity at a critical value of the non-Hermiticity parameter. This work enriches the analytical toolkit for nonadiabatic dynamics in multi-level non-Hermitian systems and provides quantitative, testable predictions for defect freezing across higher-order EPs, possibly accessible on platforms such as electric circuit networks and photonic lattices.

\end{abstract}

\maketitle

\section{Introduction}

Non-Hermitian physics \cite{Ashida-2020} is a rapidly growing area that hosts a variety of exotic phenomena without Hermitian counterparts, such as the non-Hermitian skin effect \cite{HN-1996,Zhang-2020,Okuma-2020}, exceptional points \cite{Bergholtz-2021,Ding-2022}, and anomalous bulk-edge correspondence \cite{Lee-2016,Yao-2018}. The current work focuses on theoretical study of nonequilibrium non-Hermitian quantum systems subject to time-dependent drives. For such systems, obtaining analytical solutions for their dynamics is highly desirable. 
However, since a time-dependent Hamiltonian at different instants generally does not commute, exact solution of the eigenvalues and eigenstates at any instantaneous time is insufficient to determine the dynamical evolution \cite{note-stationary}. Consequently, exactly solving a time-dependent Hermitian problem is often considered a formidable task even for two-level systems; and the situation is even more challenging for a non-Hermitian problem due to the additional complexity introduced by non-Hermiticity.

Despite these difficulties, exact solvability has been achieved for a number of time-dependent quantum systems. For two-level Hermitian models, the most famous solvable example is probably the Landau-Zener (LZ) model \cite{landau,zener,stuckelberg,majorana,Shevchenko-2010,Shevchenko-2023} with a linearly time-dependent drive. Other solvable cases include the Rosen-Zener model \cite{Rosen-1932}, the Allen-Eberly model \cite{Hioe-1984,Allen-1975}, and the Demkov-Kunike model \cite{DK-1969,Chen-2024}, which reduces to the three aforementioned models in specific limits. 
Solvable two-level non-Hermitian models, as extensions of their Hermitian counterparts, e.g. generalizations of the LZ model \cite{Torosov-2017,Shen-2019,Longstaff-2019,Sim-2023,Malla-2023,Wang-2022,Pan-2024,Hu-2025} and the Rosen-Zener model \cite{Luo-2017,Liu-2024}, have also been identified. Going beyond two levels, various Hermitian multistate generalizations of the LZ model have been solved \cite{DO,Hioe-1987,bow-tie,Rau-1998,GBT-Demkov-2000,GBT-Demkov-2001,chain-2002,4-state-2002,Rau-2003,Rau-2005,Vasilev-2007,chain-2013,Fai-2014,Patra-2015,6-state-2015,4-state-2015,DTCM-2016,DTCM-2016-2,HC-2017%
,cross-2017,quest-2017,commute,Yuzbashyan-2018,large-2018,MTLZ-2020,parallel-2020,quadratic-2021,Malla-2021,Malikis-2026}. Solvable multi-level non-Hermitian models have also been discovered, albeit relatively rarely \cite{Malla-2023,Fring-2019,Melanathuru-2022}. Methods for exactly solving time-dependent quantum problems include special functions \cite{zener,Rosen-1932,Hioe-1984,Chen-2024,Torosov-2017,Shen-2019,Longstaff-2019,Sim-2023,Pan-2024}, Laplace transformations \cite{majorana,DO,bow-tie,GBT-Demkov-2001}, Lie algebra and Lie group techniques \cite{majorana,Hioe-1987,Rau-1998,Rau-2003,Rau-2005,Fai-2014,Patra-2015,Melanathuru-2022,Malikis-2026}, analytical constraints \cite{HC-2017,cross-2017}, and integrability-based approaches \cite{Patra-2015,6-state-2015,4-state-2015,quest-2017,commute,Yuzbashyan-2018,large-2018,MTLZ-2020,parallel-2020,quadratic-2021,Malla-2021,Malla-2023}. Among these, Lie algebra/group techniques are particularly powerful for extending analytical results from two-level to multi-level systems, yet they are sometimes perceived as too mathematical for physicists.

In this work, we apply a specific Lie-algebra-based method, namely, the Wei-Norman approach \cite{Wei-1963,Wei-1964}(detailed in Sec. II), to show that for a spin-$S$ under a general non-Hermitian time-dependent drive, the matrix elements of its evolution operator (i.e. the scattering amplitudes) can be expressed in closed forms in terms of those of the corresponding spin-$1/2$ model via Jacobi polynomials. This approach is conceptually transparent and accessible to researchers without specialized expertise in Lie algebra. 
The result implies that any solvable two-level non-Hermitian quantum model can be used to construct a class of solvable multi-level models. As an application, we derive analytical expressions for the excitation probabilities in a class of nonequilibrium non-Hermitian lattices, namely, spin-$S$ extensions of the $\mathcal{PT}$-symmetric Su-Schrieffer-Heeger (SSH) model under quenches. We find that the phenomenon of defect freezing, i.e. the breakdown of adiabaticity \cite{Doppler-2016,Sim-2023}, occurs in momentum sectors that pass through a pair of higher-order exceptional points (EPs) during the quench. Our work offers two main contributions to the theoretical study of nonequilibrium non-Hermitian systems: a powerful tool for constructing solvable multi-level non-Hermitian models, and a quantitative description of defect freezing across higher-order EPs.

This paper is organized as follows. In Sec.~II, we discuss the Wei-Norman approach for a spin-$S$ under a non-Hermitian drive and derive analytical expressions of its evolution operator in terms of the spin-$1/2$ case. In Sec.~III, we consider spin-$S$ extensions of the $\mathcal{PT}$-symmetric SSH lattice under quenches and perform a detailed analytical investigation of defect freezing. In Sec.~IV, we present conclusions and outlooks.




\section{Wei-Norman approach for spin-$S$ under general non-Hermitian drive}

\subsection{Model: Spin-$S$ under non-Hermitian drive}
We consider an $N$-level quantum system whose state vector $\psi$ evolves under the Schr\"{o}dinger equation $i d\psi/dt=H\psi$ (we set $\hbar =1$) with a Hamiltonian:
\begin{align}\label{}
& H=\vec{X}\cdot \vec{S}= X_x S_x+ X_y S_y+ X_z S_z\nn\\
&=\frac{1}{2} ( X_-  S_+ + X_+ S_- )+X_z S_z ,
\label{eq:Hal-S-gen}
\end{align}
where $\vec{S}$ is a spin operator whose size $S$ can be any integer or half-integer (it is related to the number of levels by $N=2S+1$
), $S_x$, $S_y$ and $S_z$ are spin projection operators, $S_{\pm}=S_x\pm i S_y$, $\vec{X}$ is a $3$-dimensional vector, and $X_{\pm}=X_x\pm i X_y$. This model describes a single spin evolving under an external field $\vec{X}$. We will assume the {\it most general} form of $\vec{X}$, namely, the field components $X_x$, $X_y$ and $X_z$ can be any real or complex functions with any dependencies on time. If any of the three components are complex, the Hamiltonian is then non-Hermitian. The evolution operator corresponding to this Hamiltonian from time $t=t_i$ to $t=t_f$ is:
\begin{align}\label{eq:U}
U_{S}(t_f,t_i)=\mathcal{T} e^{-i\int_{t_i}^{t_f} H  dt},
\end{align}
where $\mathcal{T} $ is the time-ordering operator, and a subscript $S$ is added to emphasize the size of spin.

The fact that the Hamiltonian \eqref{eq:Hal-S-gen} belongs to the su(2) algebra (namely, it consists of only linear combinations of three generators of the su(2) algebra) allows great simplification of the evolution problem. For the Hermitian case when the field $\vec X$ is real, starting from Majorana's pioneering work \cite{majorana}, it was established using a variety of Lie algebra/group related techniques \cite{majorana,Wei-1963,Wei-1964,LL-1981,Hioe-1987,Dattoli-1988,Rau-1998,Pokrovsky-2004,Rau-2003,Rau-2005,Fai-2014,Patra-2015,Malikis-2026} that the evolution problem of an arbitrary spin $S$ under \eqref{eq:Hal-S-gen} can be reduced to that of a spin $1/2$ with the same field $\vec{X}$. 
For the non-Hermitian case with a complex $\vec X$, in \cite{Melanathuru-2022} Melanathuru, Malzard, and Graefe considered a specific form of \eqref{eq:Hal-S-gen} which is a multistate LZ model with anti-Hermitian couplings, and by acting group elements on SU(2) coherent states, they derived expressions of transition probabilities of the spin $S$ model in terms of those of the spin $S=1/2$ case.

Here we are going to treat the general non-Hermitian problem \eqref{eq:Hal-S-gen} by the Wei-Norman approach \cite{Wei-1963,Wei-1964,Rau-1998,Rau-2003,Rau-2005,Fai-2014,Dattoli-1988}. In \cite{Wei-1963,Wei-1964}, Wei and Norman considered evolution of an operator $\tilde U$ governed by a differential equation $d\tilde U/dt =\tilde H \tilde U$. 
They showed that its solution can be written as a finite product of exponential operators of the form $\tilde U=\exp(g_l H_1)\exp (g_2 H_2) \ldots \exp(g_n H_n)$, where $g_l $ ($l=1,2,\ldots,n$) are time-dependent scalar functions, $H_l$ are time-independent operators, and $n$ is the dimension of the Lie algebra generated by $\tilde H$ (see also \cite{Dattoli-1988} for a thorough discussion on this exponential decomposition method). The existence of such an exponential factorization can be utilized to simplify the evolution problem. Using Wei and Norman's method, in \cite{Fai-2014} Kenmoe and Fai obtained scattering amplitudes of the Hermitian case of the model \eqref{eq:Hal-S-gen} with an arbitrary $S$ in terms of scattering amplitudes of the $S=1/2$ model in closed forms via Gauss hypergeometric functions. Since in the Wei-Norman approach the operator $i\tilde H $ is not required to be Hermitian, this approach naturally extends to the non-Hermitian case.
In the next subsection, we will show that the Wei-Norman approach enables a very straightforward derivation of the evolution operator of the general non-Hermitian model \eqref{eq:Hal-S-gen} in terms of that of the $S=1/2$ model, which is 
accessible to non-experts on Lie algebra. 


\subsection{Wei-Norman approach}

We now present the Wei-Norman approach for the model \eqref{eq:Hal-S-gen}. Our goal is to express the evolution operator $U_{S}(t_f,t_i)$ for an arbitrary $S$ in terms of elements of $U_{1/2}(t_f,t_i)$, namely, the evolution operator in the $S=1/2$ case.

The approach starts from writing the evolution operator \eqref{eq:U} as a product of exponential operators. We follow Rau's choice \cite{Rau-1998} of decomposition of the evolution operator of the model \eqref{eq:Hal-S-gen}: 
\begin{align}\label{eq:U-exp}
U_S(t_f,t_i)=e^{-i\mu_+ S_+} e^{-i\mu_- S_-} e^{-i\mu_z S_z},
\end{align}
where $\mu_\pm$  and $\mu_z$ are time-dependent functions which satisfy a set of differential equations \cite{note-ODE-mu}. In \cite{Rau-1998}, Rau showed that the equation on ${\mu}_+ $ can be transformed into a second-order linear differential equation of a form equivalent to the Schr\"{o}dinger equation of the $S=1/2$ model itself; if the $S=1/2$ model is solvable, so is ${\mu}_+ $, and then $\mu_z$ and $\mu_-$ can in turn be determined. Here in our derivation the specific forms of the equations on the functions $\mu_\pm$  and $\mu_z$ are not important and we will not attempt to solve these equations; instead, we use the crucial fact that $\mu_\pm$ and $\mu_z$ are {\it the same} for any $S$, so they serve as a bridge to connect evolution operators of the general $S$ case to that of the $S=1/2$ case. Below we show explicitly how this is achieved. 

We will use eigenstates of $S_z$ as the basis of all operators, and label these states by $m$, namely, the spin projection quantum number along $z$-direction. $m$ will also be used as matrix indices which run from $S$ to $-S$ in decreasing order (this convention turns out to be more convenient compared to the usual convention of writing the indices as from $1$ to $N$). In this $S_z$-basis, the matrix elements of spin operators read: 
\begin{align}
&(S_+)_{m,m'}
=\delta_{m,m'+1}\sqrt{(S+1-m)(S+m)},\label{eq:Sp}\\
&(S_-)_{m,m'}
=\delta_{m+1,m'}\sqrt{(S+1-m')(S+m')},\label{eq:Sm}\\
&(S_z)_{m,m'}= \delta_{m,m'}m \label{eq:Sz}.
\end{align}
At $S=1/2$, using Eqs.~\eqref{eq:Sp}-\eqref{eq:Sz}, the evolution operator \eqref{eq:U-exp} can be evaluated directly: 
\begin{align}\label{}
&U_{1/2}(t_f,t_i)=e^{-i\mu_+ S_+} e^{-i\mu_- S_-} e^{-i\mu_z S_z}\nn\\
&=\left(
\begin{array}{cc}
 e^{-\frac{i  \mu_z}{2} } (1- \mu_+ \mu_-) & -i e^{\frac{i  \mu_z}{2} } \mu_+ \\
 -i e^{-\frac{i  \mu_z}{2} } \mu_- & e^{\frac{i\mu_z}{2}} \\
\end{array}
\right)
\equiv\left( \begin{array}{cc}
u_{\uparrow\uparrow}   &  u_{\uparrow\downarrow}  \\
u_{\downarrow\uparrow} &   u_{\downarrow\downarrow}
\end{array} \right),
\end{align}
where $\uparrow$ or $\downarrow$ stands for $m= 1/2$ or $m=-1/2$, respectively. We then have
\begin{align}\label{eq:mu-and-u}
& e^{-\frac{i  \mu_z}{2} } (1- \mu_+ \mu_-)=u_{\uparrow\uparrow},\nn\\
& -i e^{\frac{i  \mu_z}{2} } \mu_+ = u_{\uparrow\downarrow},\nn\\
& -i e^{-\frac{i  \mu_z}{2} } \mu_- = u_{\downarrow\uparrow},\nn\\
&e^{\frac{i\mu_z}{2}} =u_{\downarrow\downarrow},
\end{align}
from which we can solve for $\mu_{\pm}$ and $e^{i\mu_z/2}$:
\begin{align}\label{eq:mu-in-U}
& \mu_{+}= i \frac{u_{\uparrow\downarrow}}{u_{\downarrow\downarrow}}, \quad \mu_{-}= i u_{\downarrow\uparrow} u_{\downarrow\downarrow}, \quad e^{\frac{i\mu_z}{2}} =u_{\downarrow\downarrow}.
\end{align}
Note that among the four equations in \eqref{eq:mu-and-u}, only three are independent. This is because there is an additional constraint among the elements of the evolution operator $U_{1/2}(t_f,t_i)$, namely, $\det(U_{1/2})= u_{\uparrow\uparrow} u_{\downarrow\downarrow}-u_{\uparrow\downarrow} u_{\downarrow\uparrow}=1$. This constraint follows from the fact that the Hamiltonian \eqref{eq:Hal-S-gen} is traceless, since $\det(e^{-i \vec{X}\cdot \vec{S} \Delta t})=e^{\operatorname{tr}(-i \vec{X}\cdot \vec{S} \Delta t)}=1$. If the Hamiltonian is Hermitian or anti-Hermitian, there would be more constraints on the elements of $U_{1/2}(t_f,t_i)$.

We now write out explicitly the three exponential factors in $U_{S}(t_f,t_i)$ in \eqref{eq:U-exp} for a general $S$. From \eqref{eq:Sp}, the first factor $e^{-i\mu_+ S_+} $ is an upper triangular matrix, with elements given by:
\begin{align}\label{}
&(e^{-i\mu_+ S_+})_{m,m'}= \frac{(-i\mu_+)^{m-m'}}{(m-m')!}\prod_{l=S-m+1}^{S-m'}\sqrt{l(2S+1-l)}\nn\\
&=\frac{(-i\mu_+)^{m-m'}}{(m-m')!}\sqrt{\frac{(S-m')!(S+m)!}{(S-m)!(S+m')!}}.
\end{align}
Note that under the convention that factorial of a negative integer is infinite, this expression works for all elements of $e^{-i\mu_+ S_+}$. Similarly, from \eqref{eq:Sm}, the second factor $e^{-i\mu_- S_-} $ is a lower triangular matrix, with elements
\begin{align}\label{}
(e^{-i\mu_- S_-})_{m,m'} = \frac{(-i\mu_-)^{m'-m}}{(m'-m)!}\sqrt{\frac{(S-m)!(S+m')!}{(S-m')!(S+m)!}} .
\end{align}
Finally, from \eqref{eq:Sz}, the third factor $e^{-i\mu_z S_z}$ is a diagonal matrix with diagonal elements given by:
\begin{align}\label{}
(e^{-i\mu_z S_z} )_{m,m}=e^{-im\mu_z}.
\end{align}
Therefore, the elements of $U_S(t_f,t_i)$ read (below we suppress the time dependencies $(t_f,t_i)$ for brevity):
\begin{align}\label{eq:Us-sum}
&(U_S)_{m,m'}=(e^{-i\mu_+ S_+}e^{-i\mu_- S_-}e^{-i\mu_z S_z})_{m,m'}\nn\\
&= \sum_{l=-S}^{S}  (e^{-i\mu_+ S_+})_{m,l} (e^{-i\mu_- S_-})_{l,m'}e^{-im'\mu_z}\nn\\
&=  (-i\mu_+)^{m-S } (-i\mu_-)^{m'-S }  e^{-im'\mu_z}\nn\\
&\times\sqrt{\frac{(S+m)!(S+m')!}{(S-m)!(S-m')!}}\sum_{l=-S}^{S}\frac{(-\mu_+ \mu_-)^{S -l}(S-l)!}{(m-l)!(m'-l)!(S+l)!}.
\end{align}
Note that the range of summation in the last line is effectively from $-S$ to $\min(m,m')$. This sum can be written in terms of Gauss hypergeometric functions 
with finite number of terms, or, more conveniently, in terms of Jacobi Polynomials $P_{n}^{(\alpha,\beta)}(z)$ defined by \cite{Abramowitz-Stegun}
\begin{align}\label{eq:Jacobi}
&P_{n}^{(\alpha,\beta)}(z)=\frac{\Gamma(\alpha+n+1)}{n!\Gamma(\alpha+\beta+n+1)}\nn\\
&\times\sum_{l=0}^n{n\choose l}\frac{\Gamma(\alpha+\beta+n+l+1)}{\Gamma(\alpha+l+1)} \left(\frac{z-1}{2} \right)^l,
\end{align}
where $\Gamma(z)$ is the Gamma function, and ${n\choose l}\equiv n!/[l!(n-l)!]$ is the binomial coefficient. 
If $\alpha$ and $\beta$ are integers (which is true for the case considered here), \eqref{eq:Jacobi} can also be written as
\begin{align}\label{eq:Jacobi-integer}
&P_{n}^{(\alpha,\beta)}(z)=\frac{(\alpha+n)!}{(\alpha+\beta+n)!}\nn\\
&\times\sum_{l=0}^n  \frac{(\alpha+\beta+n+l)!}{l!(n-l)! (\alpha+ l)!} \left(\frac{z-1}{2} \right)^l.
\end{align}
Comparing the sums in \eqref{eq:Us-sum} and in \eqref{eq:Jacobi-integer}, we arrive at an expression of elements of $U_S$ in terms elements of $U_{1/2}$ which achieves our goal:
\begin{align}\label{eq:Us-result}
&(U_S)_{m,m'} =\sqrt{\frac{(S+m)!(S-m)!}{(S+m')!(S-m')!}} (u_{\uparrow\downarrow} )^{m-m'}  (u_{\uparrow\uparrow})^{m+m'}  \nn\\
&\times P_{S-m}^{(m-m',m+m')}(2 u_{\uparrow\uparrow}u_{\downarrow\downarrow}-1 ),
\end{align}
where we used \eqref{eq:mu-in-U} to express $\mu_\pm$ and $e^{i\mu_z/2}$ in terms of elements of $U_{1/2}$, and also used the identity $u_{\uparrow\uparrow} u_{\downarrow\downarrow}-u_{\uparrow\downarrow} u_{\downarrow\uparrow}=1$. Eq.~\eqref{eq:Us-result} shows that $U_S$ can be obtained once $U_{1/2}$ is known. Note that this works for the Hamiltonian \eqref{eq:Hal-S-gen} with a general complex time-dependent $\vec X$, and for evolution during any time intervals. In the next subsection we will reexpress \eqref{eq:Us-result} in another form that better illustrates the structure of $U_S$.

\subsection{Discussions}

First, we make a remark on the method of derivation of \eqref{eq:Us-result}. At first sight, this derivation seems to involve only elementary mathematics, e.g. solving simple algebraic equations and performing matrix multiplications (except for the usage of definition of Jacobi polynomials). 
But note that Lie algebra is actually used implicitly at the very first step when writing out Eq.~\eqref{eq:U-exp}, namely, the exponential decomposition of $U_S$. 
In other words, in this Wei-Norman approach, all knowledge of Lie algebra is ``encapsulated'' in the proof of validity of the formula \eqref{eq:U-exp}. (A concise derivation of Eq.~\eqref{eq:U-exp} via the Baker-Campbell-Hausdorff formula is actually available, for example, see the Appendix of \cite{Rau-2005}; but a mathematically rigorous proof demands more effort \cite{Wei-1963,Wei-1964}.) This illustrates one advantage of the current approach---compared to other approaches where Lie algebra/group techniques are used explicitly and possibly scattered in the derivation, here the derivation 
is transparent and straightforward, and accessible to non-experts on Lie algebra.

Next, we note that the result \eqref{eq:Us-result} means that if any two-level non-Hermitian quantum model is found to be exactly solvable, so is its corresponding spin-$S$ extension, namely, one will obtain a class of solvable multi-level models. This is because although the spin Hamiltonian \eqref{eq:Hal-S-gen} we considered is traceless, if we add to it a term $X_0I$ where $X_0$ is a complex function of time (a c-number) and $I_N$ is the $N\times N$ identity matrix, the only change of the evolution operator is a global factor $e^{-i\int X_0 dt}$. For $S=1/2$ (i.e. $N=2$), after such an addition, \eqref{eq:Hal-S-gen} takes the most general form of a two-level Hamiltonian. In other words, any two-level Hamiltonian can be transformed to the traceless form \eqref{eq:Hal-S-gen} with no essential change of the evolution problem. 

We now explore the result \eqref{eq:Us-result} itself. It will be helpful to write out explicitly $U_S$ at several smallest $S$ (from $S=1/2$ to $S=2$):
\begin{widetext}
\begin{align}\label{eq:Us-small}
&U_{1/2}
=\left( \begin{array}{cc}
u_{\uparrow\uparrow}   &  u_{\uparrow\downarrow}  \\
u_{\downarrow\uparrow} &   u_{\downarrow\downarrow}
\end{array} \right),\nn\quad
U_{1}
=\left( \begin{array}{ccc}
u_{\uparrow\uparrow}^2   &  \sqrt{2} u_{\uparrow\uparrow} u_{\uparrow\downarrow}   &    u_{\uparrow\downarrow}^2  \\
\sqrt{2} u_{\uparrow\uparrow} u_{\downarrow\uparrow} &  2u_{\uparrow\uparrow} u_{\downarrow\downarrow}-1 &  \sqrt{2} u_{\downarrow\downarrow} u_{\uparrow\downarrow} \\
  u_{\downarrow\uparrow}^2  &  \sqrt{2} u_{\downarrow\downarrow} u_{\downarrow\uparrow}   &  u_{\downarrow\downarrow}^2
\end{array} \right),\nn\\
&U_{3/2}
=\left( \begin{array}{cccc}
u_{\uparrow\uparrow}^3   &  \sqrt{3} u_{\uparrow\uparrow}^2 u_{\uparrow\downarrow}     &   \sqrt{3} u_{\uparrow\uparrow}u_{\uparrow\downarrow}^2        & u_{\uparrow\downarrow}^3  \\
\sqrt{3} u_{\uparrow\uparrow}^2 u_{\downarrow\uparrow}  & u_{\uparrow\uparrow}(3u_{\uparrow\uparrow}u_{\downarrow\downarrow}-2)  &  u_{\uparrow\downarrow}(3u_{\uparrow\uparrow}u_{\downarrow\downarrow}-1)    &   \sqrt{3} u_{\downarrow\downarrow}u_{\uparrow\downarrow}^2  \\
\sqrt{3} u_{\uparrow\uparrow} u_{\downarrow\uparrow}^2  &  u_{\downarrow\uparrow}(3u_{\uparrow\uparrow}u_{\downarrow\downarrow}-1)     &  u_{\downarrow\downarrow}(3u_{\uparrow\uparrow}u_{\downarrow\downarrow}-2) &   \sqrt{3} u_{\downarrow\downarrow}^2 u_{\uparrow\downarrow} \\
u_{\downarrow\uparrow}^3  & \sqrt{3} u_{\downarrow\downarrow}u_{\downarrow\uparrow}^2   &  \sqrt{3} u_{\downarrow\downarrow}^2u_{\downarrow\uparrow}    &  u_{\downarrow\downarrow}^3
\end{array} \right),\nn\\
&U_{2}
=\left( \begin{array}{ccccc}
u_{\uparrow\uparrow}^4   &  2 u_{\uparrow\uparrow}^3 u_{\uparrow\downarrow}     &   \sqrt{6} u_{\uparrow\uparrow}^2u_{\uparrow\downarrow}^2        &   2 u_{\uparrow\uparrow} u_{\uparrow\downarrow}^3   & u_{\uparrow\downarrow}^4  \\
2 u_{\uparrow\uparrow}^3 u_{\downarrow\uparrow}  & u_{\uparrow\uparrow}^2(4u_{\uparrow\uparrow}u_{\downarrow\downarrow}-3)  &  \sqrt{6} u_{\uparrow\uparrow} u_{\uparrow\downarrow}(2u_{\uparrow\uparrow}u_{\downarrow\downarrow}-1) & u_{\uparrow\downarrow}^2(4u_{\uparrow\uparrow}u_{\downarrow\downarrow}-1)    &  2 u_{\downarrow\downarrow}u_{\uparrow\downarrow}^3   \\
\sqrt{6} u_{\uparrow\uparrow}^2 u_{\downarrow\uparrow}^2  &  \sqrt{6} u_{\uparrow\uparrow} u_{\downarrow\uparrow}(2u_{\uparrow\uparrow}u_{\downarrow\downarrow}-1) & \frac{3}{2}(2u_{\uparrow\uparrow}u_{\downarrow\downarrow}-1)^2-\frac{1}{2} & \sqrt{6} u_{\downarrow\downarrow} u_{\uparrow\downarrow}(2u_{\uparrow\uparrow}u_{\downarrow\downarrow}-1) &  \sqrt{6} u_{\downarrow\downarrow}^2 u_{\uparrow\downarrow}^2 \\
2 u_{\uparrow\uparrow} u_{\downarrow\uparrow}^3  &  u_{\downarrow\uparrow}^2(4u_{\uparrow\uparrow}u_{\downarrow\downarrow}-1)  & \sqrt{6} u_{\downarrow\downarrow} u_{\downarrow\uparrow}(2u_{\uparrow\uparrow}u_{\downarrow\downarrow}-1) & u_{\downarrow\downarrow}^2(4u_{\uparrow\uparrow}u_{\downarrow\downarrow}-3)  &   2 u_{\downarrow\downarrow}^3 u_{\uparrow\downarrow}   \\
u_{\downarrow\uparrow}^4  & 2 u_{\downarrow\downarrow}u_{\downarrow\uparrow}^3   &   \sqrt{6} u_{\downarrow\downarrow}^2u_{\downarrow\uparrow}^2 & 2 u_{\downarrow\downarrow}^3u_{\downarrow\uparrow}    &  u_{\downarrow\downarrow}^4
\end{array} \right).
\end{align}
\end{widetext}
One can see from these explicit matrices that $U_{S}$ possesses certain reflection symmetries
. Indeed, using properties of Jacobi polynomials, one can recast \eqref{eq:Us-result} into a form that better illustrates these symmetries:
\begin{align}\label{eq:Us-result-2}
&(U_S)_{m,m'} = \sqrt{\frac{(S+\overline{m})!(S-\overline{m})!}{(S+\underline{m})!(S-\underline{m})!}}(u_1 )^{|m-m'|}  (u_2)^{|m+m'|}  \nn\\
&\times P_{S-\overline{m}}^{(|m-m'|,|m+m'|)}(2 u_{\uparrow\uparrow}u_{\downarrow\downarrow}-1 ),
\end{align}
where for notation simplicity we denoted $\overline{m}\equiv\max(|m|,|m'|)$, $\underline{m}\equiv\min(|m|,|m'|)$, and
\begin{align}\label{}
&u_1= \left\{  \begin{array}{l}
 u_{\uparrow\downarrow} ,\textrm{ for } m \ge m' , \\
  u_{\downarrow\uparrow} ,\textrm{ for } m \le m' ,
  \end{array}\right.\\
&u_2= \left\{  \begin{array}{l}
 u_{\uparrow\uparrow} ,\textrm{ for } m\ge-m' , \\
  u_{\downarrow\downarrow} ,\textrm{ for } m\le-m' .
  \end{array}\right.
\end{align}
Each element $(U_{S})_{m,m'}$ in \eqref{eq:Us-result-2} includes a non-negative power of $u_{\ua\da}$ or $u_{\da\ua}$, a non-negative power of $u_{\ua\ua}$ or $u_{\da\da}$, and a Jacobi polynomial $P_{n}^{(\alpha,\beta)}(z)$ with  non-negative $\alpha $ and $\beta $ (in contrast, in \eqref{eq:Us-result} the powers of $u_{\ua\da}$ and $u_{\ua\ua}$ may be negative, and $\alpha $ and $\beta$ in the Jacobian polynomial may be negative). In particular, the element $(U_S)_{m,m'}$ contains a factor of power of $u_{\uparrow\downarrow} $ or $u_{\downarrow\uparrow} $ if it is closer to the top-right corner or to the bottom-left corner, respectively; it contains a factor of power of $u_{\uparrow\uparrow} $ or $u_{\downarrow\downarrow} $ if it is closer to the top-left corner or to the bottom-right corner, respectively. The powers depend on how close the element is to the relevant corners. 
The reflection symmetries of the matrix $U_S$ can be stated as follows: first, $U_S$ is invariant upon sending $m\leftrightarrow m'$ (a reflection with respect to the main diagonal) and $u_{\uparrow\downarrow}\leftrightarrow u_{\downarrow\uparrow} $; second, it is also invariant upon sending $m\leftrightarrow -m'$ (a reflection with respect to the antidiagonal) and $u_{\uparrow\uparrow}\leftrightarrow u_{\downarrow\downarrow} $.



From \eqref{eq:Us-small}, one can also observe that elements at the edges of the matrix, namely, $(U_S)_{m,m'}$ where one of $m$ or $m'$ is $\pm S$, allow simple expressions. 
For example, for $m'=S$, namely for elements in the leftmost column of $U_S$, we have from \eqref{eq:Us-result-2}
\begin{align}\label{eq:U-S-j1}
&(U_S)_{m,S}  =\sqrt{\frac{(2S)!}{(S+m)!(S-m)!}}(u_{\downarrow\uparrow} )^{S-m}  (u_{\uparrow\uparrow})^{S+m}\nn\\
&\times P_{0}^{(S-m,S+m)}(2 u_{\uparrow\uparrow}u_{\downarrow\downarrow}-1 )\nn\\
&=\sqrt{2S\choose S+m} (u_{\downarrow\uparrow} )^{S-m}  (u_{\uparrow\uparrow})^{S+m} .
\end{align}
This expression will be useful in the next section.

Finally, we note that the result \eqref{eq:Us-result} for the general non-Hermitian case is consistent with previous results for the Hermitian case (namely, when the field $\vec X$ is real). In the latter case $U_{1/2}$ is unitary, and we have the constraints $u_{\uparrow\downarrow} = - u_{\downarrow\uparrow}^* $ and $u_{\uparrow\uparrow} =  u_{\downarrow\downarrow}^* $. Then one can check that the expression \eqref{eq:Us-result} reduces to the result in the Hermitian case, namely, the Wigner d-matrix 
(for example, see Eq.~(58.10) in Landau and Liftshitz's book \cite{LL-1981}).

\section{Defect freezing in spin-$S$ $\mathcal{PT}$-SSH models under quenches}

As mentioned in the introduction, non-Hermitian systems host phenomena absent in Hermitian systems, and here we are interested in a phenomenon specific to nonequilibrium non-Hermitian systems. It was experimentally observed \cite{Doppler-2016} that for a non-Hermitian system under a time-dependent drive, a breakdown of adiabaticity can take place, namely, excitations can exist even in the adiabatic limit when the drive is infinitely slow. This is in sharp contrast to the case of nonequilibrium dynamics of a Hermitian system, where the adiabatic theorem guarantees no excitations in the adiabatic limit. This interesting phenomenon was termed {\it defect freezing} \cite{Sim-2023}, and it receives considerable interest in theoretical studies \cite{Longstaff-2019,Melanathuru-2022,Wang-2022,Sim-2023,Deng-2025,Hu-2025}. In this section, we will apply the general results in the previous section to study defect freezing in a specific class of non-Hermitian systems. We construct a family of lattices 
that realize a particular spin model \eqref{eq:Hal-S-gen} in momentum space, and then evaluate analytically the amount of excitations for these models under quenches. 


\subsection{The $\mathcal{PT}$-SSH model and its spin-$S$ extensions}

\begin{figure}[!htb]
\scalebox{0.4}[0.4]{\includegraphics{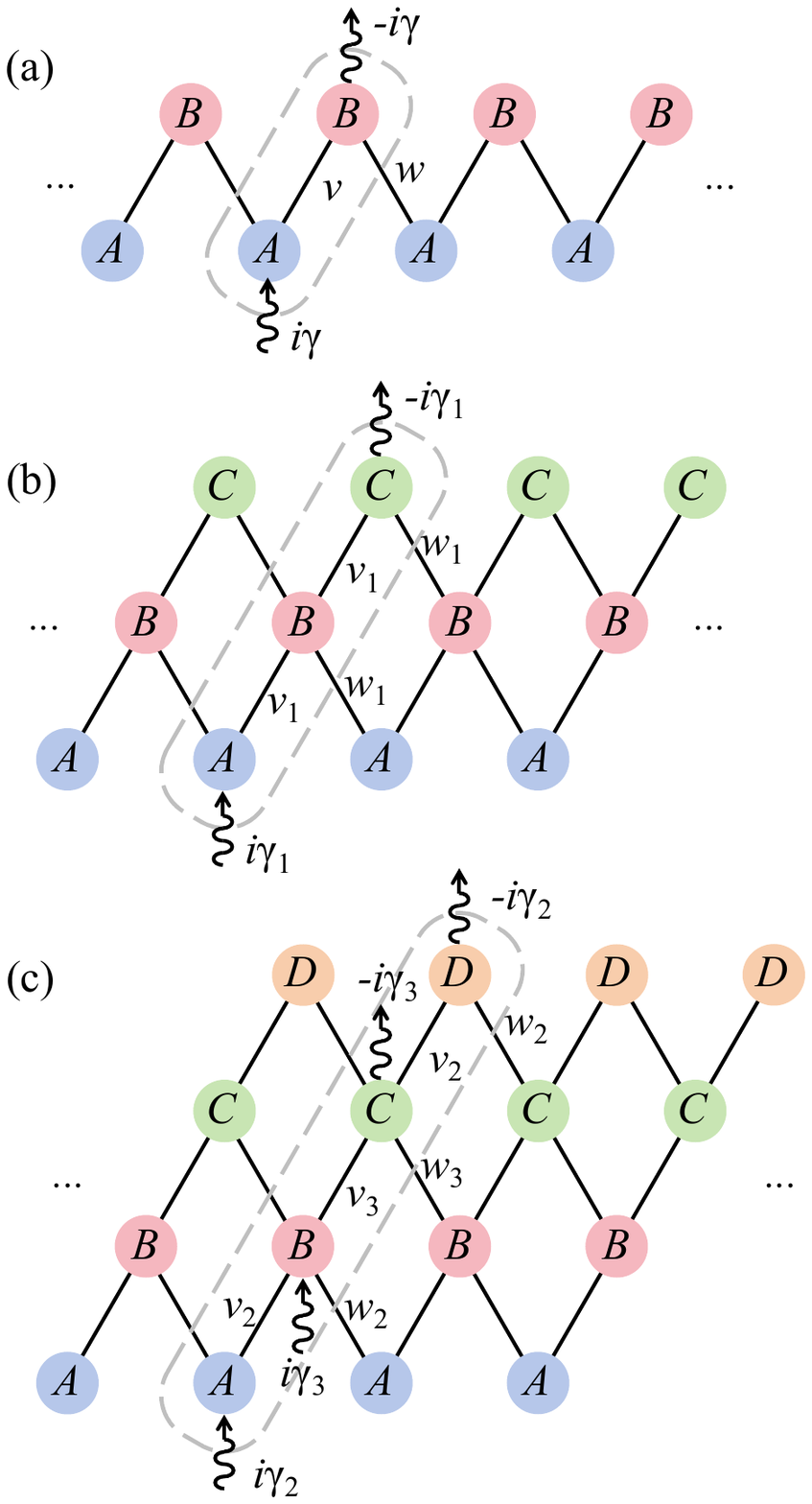}}
\caption{Sketches of lattices of (a) the $\mathcal{PT}$-SSH model, (b) its $S=1$ extension, and (c) its $S=3/2$ extension. $A$, $B$, $C$ and $D$ stand for sites in different sublattices, a solid line presents intracell or intercell hoppings between sites, and a waved line presents gain or loss at a given site. In each sketch four unit cells are plotted, with a dashed line encircling a specific unit cell, and only parameters within this cell and between it and its adjacent cell to the right are shown (parameters in the whole lattice can be read off by translational symmetry). The parameters of the $S=1$ and $S=3/2$ models are related to those of the $S=1/2$ model as: $v_1=\sqrt 2 v$, $w_1=\sqrt 2 w$, $\gamma_1=2 \gamma$, $v_2=\sqrt 3 v$, $v_3=2 v$, $w_2=\sqrt 3 w$, $w_3=2 w$, $\gamma_2=3\gamma$, $\gamma_3=\gamma$. }
\label{fig:nSSH}
\end{figure}

We consider a class of models that builds on a non-Hermitian lattice model, namely, the $\mathcal{PT}$-symmetric Su-Schrieffer-Heeger ($\mathcal{PT}$-SSH) model \cite{Lieu-2018,Gong-2018,Deng-2025,Zhou-2026}. The original SSH model \cite{SSH1,SSH2,Asboth-2016} is a diatomic chain with different intracell and intercell hopping amplitudes. Adding an imaginary staggered potential to each site
, one obtains the $\mathcal{PT}$-SSH model, as sketched in Fig.~\ref{fig:nSSH}(a). Under periodic boundary conditions, the real-space Hamiltonian of this model can be Fourier transformed into a block-diagonal form with each $2\times 2$ block corresponding to a momentum $k$ \cite{Lieu-2018}. States in the sector with momentum $k$ then evolves under a Shr\"{o}dinger equation $i d\psi/dt=H\psi$ with a Hamiltonian:
\begin{align}\label{}
& H=\left(\begin{array}{cc}
 i\gamma & v+we^{-ik} \\
 v+we^{ik}  & -i\gamma
\end{array}\right)\nn\\
&= (v+w\cos k) \sigma_x+w \sin k  \sigma_y+ i\gamma \sigma_z,
\label{eq:n-SSH}
\end{align}
where $v$ and $w$ are the intracell and intercell hoppings, respectively, $\pm i\gamma$ is the on-site imaginary staggered potential (for $\gamma>0$, $i\gamma$ and $-i\gamma$ correspond to gain and loss, respectively)
, and $\sigma_{x/y/z}$ are the Pauli matrices. Note that $|\gamma|$ represents the strength of non-Hermiticity; at $\gamma=0$ the original (Hermitian) SSH model is recovered. We denote the total number of unit cells in the whole chain as $L$. When $L$ is large, $k$ can be treated as a continuous variable with a range $k\in (-\pi,\pi]$. 
The Hamiltonian \eqref{eq:n-SSH} possesses $\mathcal{PT}$ symmetry \cite{Bender-1998} in the sense that $[H,\mathcal{PT}]=0$, with the parity operator $\mathcal{P} =\sigma_z$ and the time-reversal operator $\mathcal{T}=-i\sigma_y K$ where $K$ is complex conjugation. This symmetry guarantees that the spectrum of \eqref{eq:n-SSH} is real in some regions in the parameter space, but there can also be regions with a complex spectrum, where the $\mathcal{PT}$ symmetry is said to be spontaneously broken. Specifically, the spectrum of \eqref{eq:n-SSH} reads: $E_{k,\pm}=\pm\sqrt{v^2+w^2+2vw\cos k-\gamma^2}$. When $|\gamma|<|v-w|$, $E_{k,\pm}$ are real for all $k$, and all $k$ sectors are $\mathcal{PT}$-symmetric. On the other hand, when $|\gamma|>|v-w|$, $E_{k,\pm}$ are purely imaginary for some $k$, and these $k$ sectors are $\mathcal{PT}$-broken. 
In the critical case $v^2+w^2+2vw\cos k-\gamma^2=0$, we have $E_{k,+}=E_{k,-}=0$, and the Hamiltonian \eqref{eq:n-SSH} is at a second order exceptional point (EP) where its two eigenstates coalesce. In the parameter space, such an EP lies on the boundaries separating the $\mathcal{PT}$-symmetric and $\mathcal{PT}$-broken regions.

We now rewrite the momentum-space Hamiltonian \eqref{eq:n-SSH} in terms of spin operators (note that $\vec{S}=\vec{\sigma}/2$ at $S=1/2$):
\begin{align}\label{}
&H=2 [(v+w\cos k) S_x+w \sin k  S_y+ i\gamma S_z].
\label{eq:n-SSH-S}
\end{align}
Treating the spin operators in \eqref{eq:n-SSH-S} as those of a general spin $S$, we obtain an $N\times N$ momentum-space Hamiltonian (recall that $N=2S+1$) of the form \eqref{eq:Hal-S-gen}. For example, for $S=1$, one obtains a $3\times3$ Hamiltonian:
\begin{align}\label{}
&H=\left( \begin{array}{ccc}
2i\gamma & \sqrt{2}(v+we^{-ik}) &  0\\
\sqrt{2}( v+we^{ik}) & 0 &   \sqrt{2}(v+we^{-ik})\\
0 & \sqrt{2}(v+we^{ik})  & -2i\gamma
\end{array} \right).
\label{eq:n-SSH-S=1}
\end{align}
The spectrum of \eqref{eq:n-SSH-S} with spin $S$ reads: $E_{k,m}=2m\sqrt{v^2+w^2+2vw\cos k-\gamma^2}$, where $m$ ranges from $S$ to $-S$. Therefore, the conditions of the $\mathcal{PT}$-symmetric and $\mathcal{PT}$-broken regions are the same as those for the $S=1/2$ case. At $v^2+w^2+2vw\cos k-\gamma^2=0$, we have $E_{k,m}=0$ for all $m$, and the Hamiltonian \eqref{eq:n-SSH-S} is at an $N$th order EP.

The $N\times N$ momentum-space Hamiltonian \eqref{eq:n-SSH-S} can be realized by a lattice that extends the $\mathcal{PT}$-SSH chain in Fig.~\ref{fig:nSSH}(a) to include $N$ sites per unit cell. In Fig.~\ref{fig:nSSH}(b) and (c), we sketch these lattices at $S=1$ and $S=3/2$, respectively, whose parameters are determined such that the corresponding momentum space Hamiltonian is \eqref{eq:n-SSH-S}. 
Extensions to larger spins follow the same manner. Interestingly, as the spin becomes larger, the lattice grows from a one-dimensional chain to a two-dimensional array. 
We call this family of models {\it spin-$S$ $\mathcal{PT}$-SSH models}. We expect that these models can be realized experimentally on various platforms \cite{Weimann-2017,Stegmaier-2021,Qian-2024}; possible experimental realizations will be discussed in more detail at the end of this section.

\subsection{Defect freezing}

In this subsection, we examine quantitatively the amount of excitations in the adiabatic limit in a nonequilibrium (time-dependent) version of the spin-$S$ $\mathcal{PT}$-SSH model \eqref{eq:n-SSH-S} with a general $S$. Specifically, we consider a linear quench of the parameter $v$ as $v=b t$ at a quench rate $b>0$ from some large negative time to some large positive time (effectively, from $t=-\infty$ to $t=\infty$). Other two parameters $w$ and $\gamma$ are set to constants, and we assume $w>0$ and $\gamma>0$. We take the initial state in each $k$ sector to be its ground state, namely, the eigenstate with the lowest energy in the $\mathcal{PT}$-symmetric region at $t=-\infty$, which is the $m=S$ diabatic state. We address two questions in the adiabatic limit $b\rar0$: first, what is the probability of excitations from the ground state (namely, the probability that a defect is generated) for each $k$ sector after the quench; second, what is the total excitation density after summing the contributions from all $k$ sectors. It turns out that with Eq.~\eqref{eq:Us-result} and the exact solution of the evolution problem at $S=1/2$, analytical results can be obtained for both questions.

\subsubsection{Defect freezing in the $\mathcal{PT}$-SSH model}
We first discuss the $S=1/2$ case, namely, the original $\mathcal{PT}$-SSH model \eqref{eq:n-SSH}. In this case, the quench problem is the same as protocol III considered in \cite{Deng-2025}, except that here the evolution is between infinite times. The quench starts and ends where all $k$ sectors are $\mathcal{PT}$-symmetric, but it traverses a region where some $k$ sectors may become $\mathcal{PT}$-broken. In particular, at $v=w\cos k\pm\sqrt{\gamma^2-w^2\sin^2k}$ we have $E_{k,+}=E_{k,-}=0$, and the Hamiltonian \eqref{eq:n-SSH} is at a second order EP. Therefore, for those $k$ sectors with $|\sin k|<\gamma/w$, the quench passes through two EPs successively, between which (namely, when $w\cos k-\sqrt{\gamma^2-w^2\sin^2k}<v<w\cos k+\sqrt{\gamma^2-w^2\sin^2k}$) the sectors are $\mathcal{PT}$-broken. On the other hand, for those $k$ sectors with $|\sin k|>\gamma/w$, $E_{k,\pm}=0$ has no solutions at real $v$; these sectors always have real $E_{k,\pm}$, and they stay $\mathcal{PT}$-symmetric during the whole quench process. Thus, $\gamma=w$ is a critical situation: at $\gamma<w$ some of the $k$ sectors pass through two EPs, whereas at $\gamma>w$ all $k$ sectors pass through two EPs. Below we will see the consequences of this fact on defect freezing.

To analyze the $\mathcal{PT}$-SSH model \eqref{eq:n-SSH} under the quench, it is convenient to make a rotation of axes: $\sigma_x \leftrightarrow \sigma_z$ and $\sigma_y\rar-\sigma_y$, so the Hamiltonian becomes
\begin{align}\label{}
H
= (v+w\cos k) \sigma_z-w \sin k  \sigma_y+ i\gamma \sigma_x.
\end{align}
The linearly time-dependent term then moves into the diagonal part. Plugging in $v=bt$, one can identify this Hamiltonian as a non-Hermitian LZ model
\begin{align}\label{}
 H
 =  \left( \begin{array}{cc}
 b (t-t_0)  &  g_{12}  \\
g_{21} &  -b(t-t_0)
\end{array} \right)
\label{eq:nLZ}
\end{align}
with parameters
\begin{align}\label{eq:nLZ-para}
& t_0= -\frac{w}{b}\cos k,\quad g_{12}=i(\gamma+w\sin k),\nn\\
& g_{21}=i(\gamma-w\sin k).
\end{align}
The non-Hermitian LZ model can be exactly solved by a special function approach similar to one for a Hermitian LZ model. Shen et~al. \cite{Shen-2019} performed a thorough investigation on the solution of the non-Hermitian LZ model \eqref{eq:nLZ} with the most general complex couplings $g_{12}$ and $g_{21}$. There were also studies focusing on specific cases of the model, for example, Torosov and Vitanov \cite{Torosov-2017} and Wang et~al. \cite{Wang-2022} considered the nonreciprocal case where $g_{12}$ and $g_{21}$ take different real values, Longstaff and Graefe \cite{Longstaff-2019} considered the anti-Hermitian case where $g_{12}=g_{21}=i\gamma$, and Pan and Wu \cite{Pan-2024} considered the $\mathcal{PT}$-symmetric case as in the present work. Here for our model \eqref{eq:nLZ} with parameters \eqref{eq:nLZ-para}, the exact analytical expression of excitation probability from the ground state after the quench (in the sector with momentum $k$) reads \cite{Deng-2025}:
\begin{align}\label{eq:pk-1o2}
&p_{k,1/2}=\frac{ \gamma+w\sin k}{2\gamma-(\gamma-w\sin k) e^{2 \pi\delta}},
\end{align}
where $\delta= ( w^2\sin^2k -\gamma^2)/(2b)$, and the subscript $1/2$ means $S=1/2$ (recall that the $S=1/2$ model is the original $\mathcal{PT}$-SSH model). We refer to Sec. V in Supplemental Material of \cite{Deng-2025} for derivation of \eqref{eq:pk-1o2}. 
Note that the result \eqref{eq:pk-1o2} is obtained by a direct normalization of the unnormalized probabilities \cite{Deng-2025}; one will get a different result if the metric formalism of time-dependent non-Hermitian problems \cite{Sim-2023,Sim-2025,Hu-2025} is adopted instead. Also note that in the Hermitian limit with $\gamma=0$, one recovers from \eqref{eq:pk-1o2} the standard result of a Hermitian LZ problem, namely, the famous LZ formula: $p_{k,1/2}= e^{-2 \pi\delta}$.

In the adiabatic limit $b\rar 0$, we see that if $\delta>0$, then $e^{2\pi \delta}\rar\infty$, which gives $p_k\rar 0$. On the other hand, if $\delta<0$, then $e^{2\pi \delta}\rar0$, and $p_k=(\gamma+w\sin k)/(2\gamma)$. Eq.~\eqref{eq:pk-1o2} then gives
\begin{align}\label{eq:pk-1o2-adiabatic}
p_{k,1/2}(b\rar 0)=\left\{\begin{array}{cc}
             \frac{1}{2}+ \frac{w\sin k}{2\gamma}, & \textrm{for }   |\sin k|<\frac{\gamma}{w}, \\
           0, & \textrm{for }  |\sin k| >\frac{\gamma}{w}.
           \end{array}
\right.
\end{align}
The non-zero value of $p_{k,1/2}(b\rar 0)$ signals defect freezing: non-zero number of defects are generated even in the adiabatic limit. Recall that the $k$ sectors with $|\sin k|<\gamma/w$ passes the $\mathcal{PT}$-broken region during the quench, whereas the $k$ sectors with $|\sin k|>\gamma/w$ always stay in the $\mathcal{PT}$-symmetric region during the quench. Therefore, Eq.~\eqref{eq:pk-1o2-adiabatic} is consistent with the general statement that the phenomenon of defect freezing is associated with evolution inside the $\mathcal{PT}$-broken region, or equivalently, evolution across EPs.

Below we consider the (total) excitation density, namely, number of excitations per unit cell. It is defined by summing $p_{k,1/2} $ over all the $k$ sectors:
\begin{align}\label{eq:nex-def}
n_{ex,1/2}=\frac{1}{L}\sum_k p_{k,1/2} \rar \frac{1}{2\pi}\int_{-\pi}^{\pi} p_{k,1/2} dk.
\end{align}
In the adiabatic limit, $n_{ex,1/2}$ has a simple analytical form:
\begin{align}\label{}
n_{ex,1/2}(b\rar 0)= \left\{\begin{array}{cc}
\frac{1}{\pi}\arcsin\frac{\gamma}{w}, & \textrm{for }  \gamma<w, \\
            \frac{1}{2},  & \textrm{for }  \gamma\ge w.
           \end{array}
\right.
\end{align}
We see that at $\gamma=0$ (the Hermitian limit), $n_{ex}$ vanishes, which is expected, since for a Hermitian system no defects are generated in the adiabatic limit. As $\gamma$ increases, $n_{ex}$ also increase; and at $\gamma=w$ it reaches $1/2$, after which it stays as $1/2$ as $\gamma$ increases further. The non-analytical behavior at $\gamma=w$ originates from the fact that from \eqref{eq:pk-1o2-adiabatic}, for $\gamma<w$, $p_{k,1/2}(b\rar 0)$ is non-zero only in the regions near $k=0$ and $k=\pi$ with $|\sin k|<\gamma/w$, and for $\gamma> w$, $p_{k,1/2}(b\rar 0)$ is non-zero at all $k$. Integration in this two situations then results in different analytical functions. Therefore, we see that the physical reason of this singularity at $\gamma=w$ is that for $\gamma< w$, some $k$ sectors do not traverse the $\mathcal{PT}$-broken region, whereas for $\gamma>w$, all the $k$ sectors traverse the $\mathcal{PT}$-broken region. 



\subsubsection{Defect freezing in the spin-$S$ $\mathcal{PT}$-SSH model}

We now consider the spin-$S$ $\mathcal{PT}$-SSH model, namely, the model \eqref{eq:n-SSH-S}, under the same quench $v=bt$ from $t=-\infty$ to $t=\infty$. This model is a non-Hermitian multistate LZ model, which describes a quench across a pair of $N$th-order EPs (recall that $N=2S+1$). Note that the anti-Hermitian model considered in \cite{Melanathuru-2022} corresponds to setting $k=0$ in the current model.

As discussed in Sec. II, solution of the spin-$1/2$ case of \eqref{eq:n-SSH} directly determines solution of the spin-$S$ case at any $S$. In particular, with the analytical expression of $p_{k,1/2}$ in \eqref{eq:pk-1o2}, we can calculate analytically the probability of excitation from the ground state for the spin-$S$ $\mathcal{PT}$-SSH model, which we denote as $p_{k,S}$.  

We first express $p_{k,S}$ in terms of $p_{k,1/2}$. At $t\rar -\infty$ the ground state is the $m'=S$ 
diabatic state. According to \eqref{eq:U-S-j1}, the unnormalized probabilities from the level $m'=S$ read to the level $m$ read:
\begin{align}\label{}
&
|(U_S)_{m,S} |^2= {2S\choose S+m}  (|u_{\downarrow\uparrow}|^2 )^{S-m}  (|u_{\uparrow\uparrow}|^2)^{S+m}.
\end{align}
At $t\rar \infty$ the ground state becomes the $m'=-S$ 
diabatic state, which would be the final state if the evolution is completely adiabatic. So the excitation probability from the ground state in the $k$ sector is
\begin{align}\label{eq:pks}
&p_{k,S}=1-\frac{|(U_S)_{-S,S}|^2 }{\sum_{m=-S}^{S}|(U_S)_{m,S} |^2}\nn\\
& =1- \left(\frac{|u_{\downarrow\uparrow}|^2}{ |u_{\uparrow\uparrow}|^2+|u_{\downarrow\uparrow}|^2}\right)^{2S}
 =1- \left(1-p_{k,1/2}\right)^{2S}.
\end{align}
Thus, $p_{k,S}$ depends solely on $p_{k,1/2}$.

\begin{figure*}[!htb]
\scalebox{0.43}[0.43]{\includegraphics{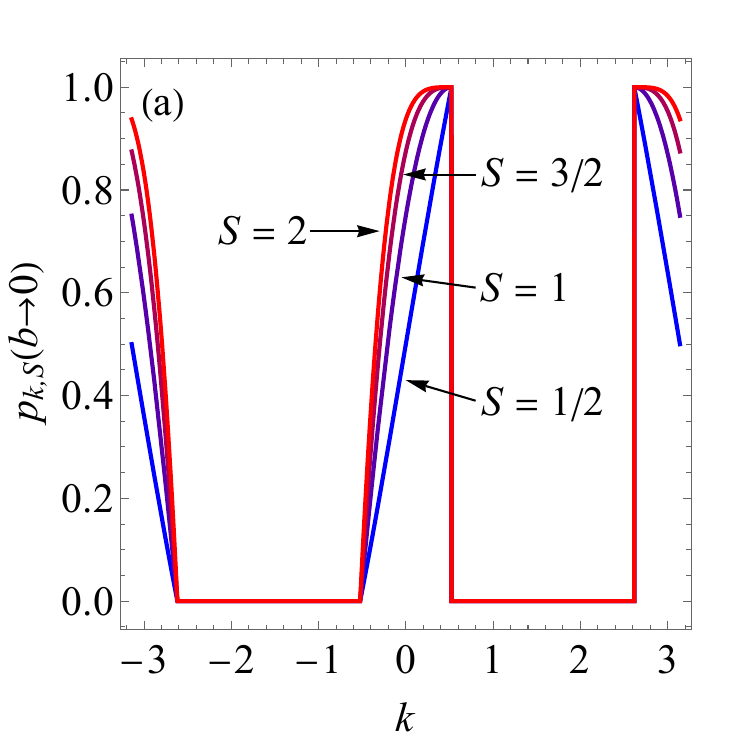}}
\scalebox{0.43}[0.43]{\includegraphics{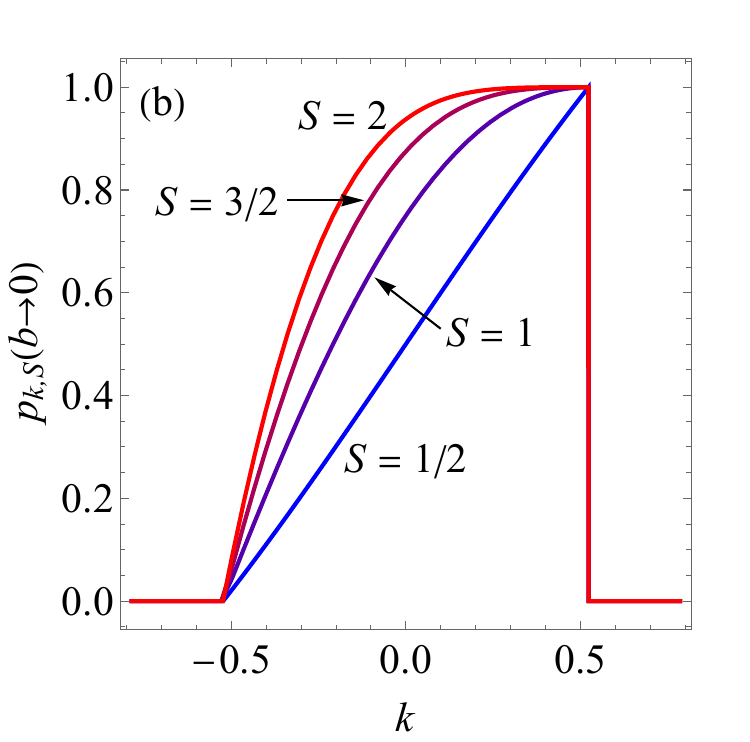}}
\scalebox{0.43}[0.43]{\includegraphics{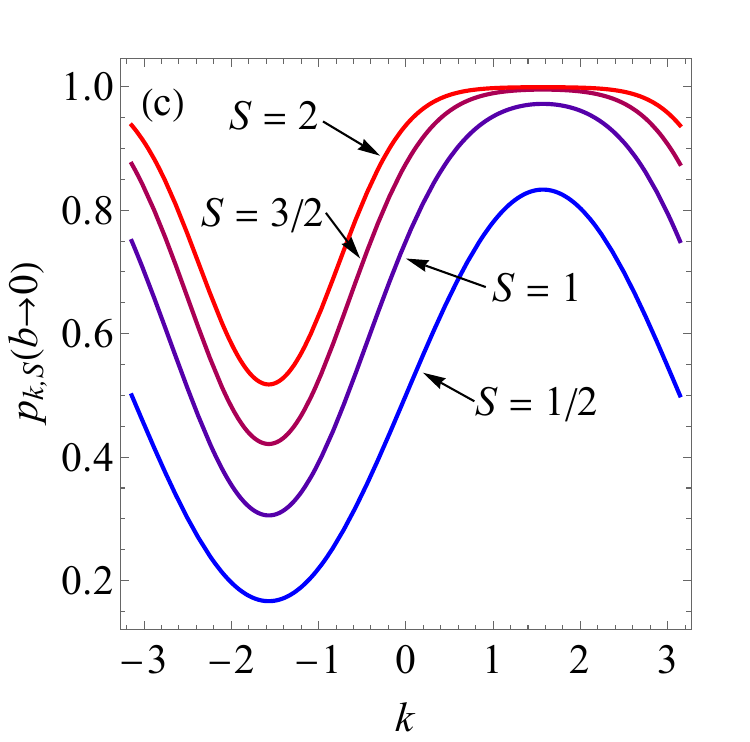}}
\caption{Excitation probability in the adiabatic limit $p_{k,S}(b\rar 0)$ vs. momentum $k$ by Eq.~\eqref{eq:pksb0} for the spin-$S$ $\mathcal{PT}$-SSH model under a quench for four different $S$ from $S=1/2$ to $S=2$. (a) at $\gamma/w=0.5$; (b) same as (a) but zoomed into a region near $k=0$; (c) at $\gamma/w=1.5$.}
\label{fig:k-frozen-defect}
\end{figure*}

We now consider defect freezing in the adiabatic limit. Plugging \eqref{eq:pk-1o2-adiabatic} into \eqref{eq:pks}, the excitation probability in the $k$ sector at $b\rar0$ reads:
\begin{align}\label{eq:pksb0}
p_{k,S}(b\rar 0)=\left\{\begin{array}{cc}
             1-\left(\frac{1}{2}- \frac{w\sin k}{2\gamma}\right)^{2S}, & \textrm{for }  |\sin k| <\frac{\gamma}{w} , \\
           0, & \textrm{for }  |\sin k| >\frac{\gamma}{w}.
           \end{array}
\right.
\end{align}
Fig.~\ref{fig:k-frozen-defect} shows $p_{k,S}(b\rar 0)$ vs. $k$ by Eq.~\eqref{eq:pksb0} for four different $S$ from $S=1/2$ to $S=2$ at $\gamma/w=0.5$ [Fig.~\ref{fig:k-frozen-defect}(a) and (b)] and at $\gamma/w=1.5$ [Fig.~\ref{fig:k-frozen-defect}(c)]. We see that like the $S=1/2$ case, for a larger $S$, defect freezing takes place only when $|\sin k| <\gamma/w$, namely, when the $k$ sector passes the $\mathcal{PT}$-broken region. At $\gamma/w=0.5$, $|\sin k| <\gamma/w$ is satisfied only in regions near $k=0$ and $k=\pi$, and $p_{k,S}(b\rar 0)$ is non-zero only in these two regions. Whereas at $\gamma/w=1.5$, $|\sin k| <\gamma/w$ is satisfied for all $k$, so $p_{k,S}(b\rar 0)$ is always non-zero. Moreover, a curve for a larger $S$ is always above a curve for a smaller $S$, meaning that as $S$ increases more defects are generated in any $k$ sector.

Below we evaluate the excitation density $n_{ex,S}$ in the adiabatic limit. $n_{ex,S}$ is defined similarly as the spin-$1/2$ case in \eqref{eq:nex-def}. Since $p_{k,S}$ depends on $k$ solely via $\sin k$, it is periodic in $k$ and symmetric under a reflection with respect to $k=\pi/2$. So $n_{ex,S}$ can also be expressed as $n_{ex,S}=1/\pi\int_{-\pi/2}^{\pi/2} p_{k,S} dk$. If $\gamma<w$,
\begin{align}\label{}
&n_{ex,S}(b\rar 0)
=\frac{1}{\pi}\int_{-\arcsin\frac{\gamma}{w}}^{\arcsin\frac{\gamma}{w}} \left[1-\left(\frac{1}{2}- \frac{w\sin k}{2\gamma}\right)^{2S}\right] dk\nn\\
&=\frac{2}{\pi}\arcsin\frac{\gamma}{w}- \frac{1}{ \pi}\int_{-\arcsin\frac{\gamma}{w}}^{\arcsin\frac{\gamma}{w}}  \left(\frac{1}{2}- \frac{w\sin k}{2\gamma}\right)^{2S}  dk.
\end{align}
The integral can be performed by making a change of variable $x=1/2- w\sin k/(2\gamma)$. Then $dx=-w\cos k/(2\gamma) dk=- \sqrt{w^2/(4\gamma^2)-(1/2-x)^2}dk$, and
\begin{align}\label{}
&\int_{-\arcsin\frac{\gamma}{w}}^{\arcsin\frac{\gamma}{w}}  \left(\frac{1}{2}- \frac{w\sin k}{2\gamma}\right)^{2S}  dk
=\int_{0}^{1} \frac{ x^{2S}}{ \sqrt{\frac{w^2}{4\gamma^2}-(\frac{1}{2}-x)^2}}  dx\nn\\
&= \frac{2}{(2S+1)\sqrt{\frac{w^2}{\gamma^2}-1}} \nn\\
&\times F_1(2S+1,\frac{1}{2},\frac{1}{2};2S+2;\frac{2}{1+\frac{w}{\gamma}},\frac{2}{1-\frac{w}{\gamma}}) ,
\end{align}
where $F_1(a,b_{1},b_{2};c;x,y)$ is the Appell hypergeometric series defined by \cite{Burchnall-1940} 
\begin{align}\label{}
 F_{1}(a,b_{1},b_{2};c;x,y)=\sum _{m,n=0}^{\infty }{\frac {(a)_{m+n}(b_{1})_{m}(b_{2})_{n}}{(c)_{m+n}\,m!\,n!}}\,x^{m}y^{n},
\end{align}
and $(q)_{n}$ is the Pochhammer symbol defined by 
\begin{align}\label{eq:Pochhammer}
(q)_n=\left\{\begin{array}{cc}
               1, & n=0, \\
               q(q+1)\cdots(q+n-1), & n>0.
             \end{array}
\right.
\end{align}
If $\gamma\ge w$,
\begin{align}\label{eq:nex-u-larger}
&n_{ex,S}(b\rar 0)
=\frac{1}{ \pi}\int_{-\frac{\pi}{2}}^{\frac{\pi}{2}} \left[1-\left(\frac{1}{2}- \frac{w\sin k}{2\gamma}\right)^{2S}\right] dk\nn\\
&=1- \frac{1}{2^{2S}}  P_{\lfloor S \rfloor}^{(0,-2S-1/2)} \left(1-\frac{2w^2}{\gamma^2}\right),
\end{align}
where 
$P_{n}^{(\alpha ,\beta )}(z)$ is the Jacobi polynomial defined in \eqref{eq:Jacobi}, and $\lfloor S \rfloor$ means the floor function of $S$ (the greatest integer less than or equal to $S$).

\begin{figure*}[!htb]
\scalebox{0.43}[0.43]{\includegraphics{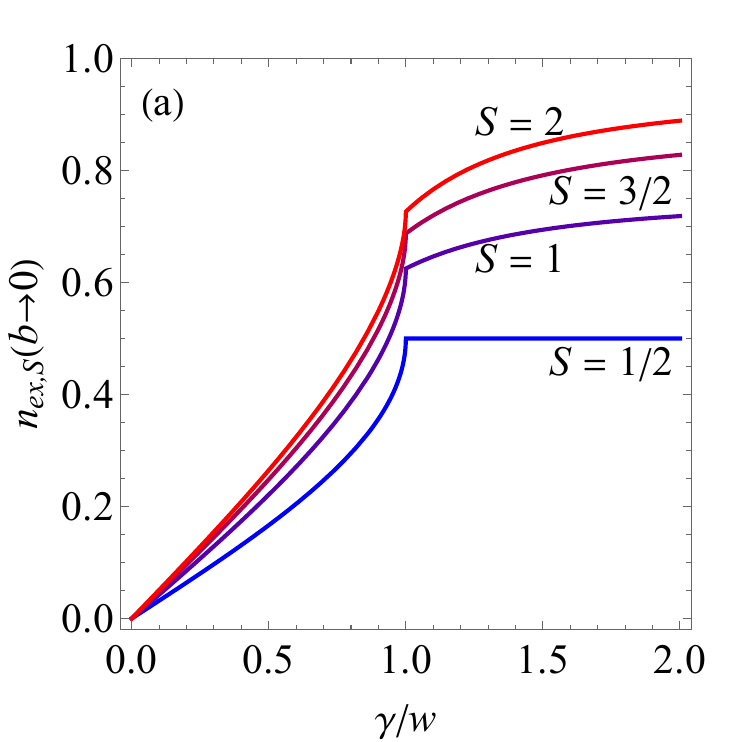}}
\scalebox{0.43}[0.43]{\includegraphics{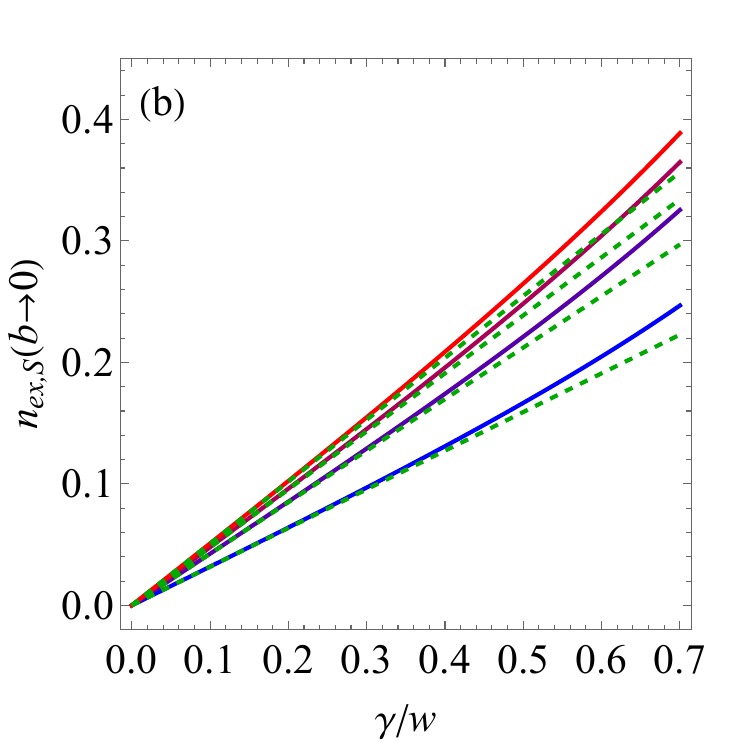}}
\scalebox{0.46}[0.46]{\includegraphics{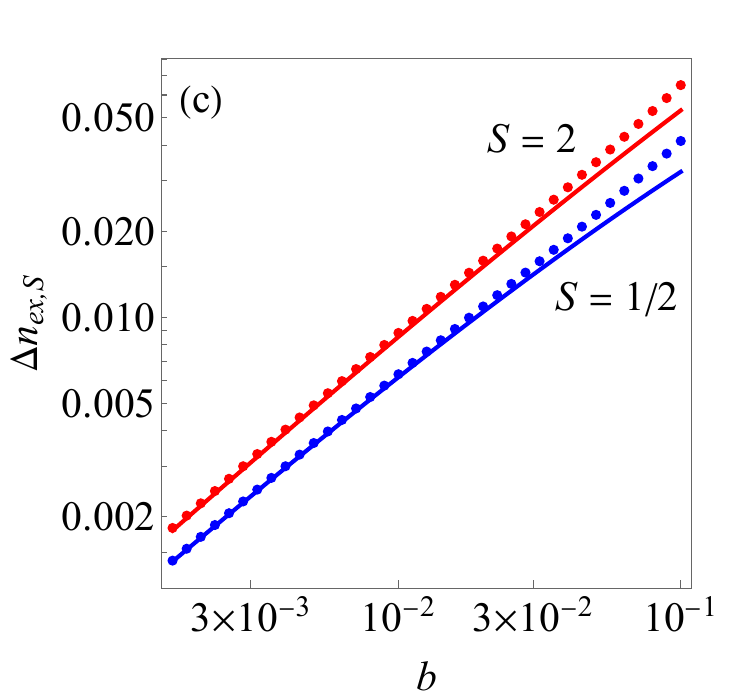}}
\caption{Excitation density in the spin-$S$ $\mathcal{PT}$-SSH model under a quench. (a) Excitation density $n_{ex,S}(b\rar 0)$ vs. $\gamma/w$ by Eq.~\eqref{eq:nex} for four different $S$ from $S=1/2$ to $S=2$. (b) The solid lines are the same as in (a) but for small $\gamma/w$, and the dashed lines are linear approximation by Eq.~\eqref{eq:nex-linear}. (c) Excitation density above the frozen defect $\Delta n_{ex,S}$ vs. quench rate $b$ at $w=1$, $\gamma=0.3$ for $S=1/2$ and $S=2$. The solid lines are by the approximate analytical expression Eq.~\eqref{eq:nex}, and the dots are from exact numerical integrations.}
\label{fig:frozen-defect}
\end{figure*}

Summarizing, the excitation density in the adiabatic limit reads:
\begin{widetext}
\begin{align}\label{eq:nex}
&n_{ex,S}(b\rar 0)=\left\{\begin{array}{cc}
             \frac{2}{\pi}\left[\arcsin\frac{\gamma}{w}-\frac{1}{(2S+1)\sqrt{\frac{w^2}{\gamma^2}-1}} F_1\left(2S+1,\frac{1}{2},\frac{1}{2};2S+2;\frac{2}{1+\frac{w}{\gamma}},\frac{2}{1-\frac{w}{\gamma}}\right)\right], & \textrm{for }   \gamma< w, \\
           1- \frac{1}{2^{2S}}  P_{\lfloor S \rfloor}^{(0,-2S-1/2)} \left(1-\frac{2w^2}{\gamma^2}\right), & \textrm{for }  \gamma\ge w.
           \end{array}\right.
\end{align}
\end{widetext}
Fig.~\ref{fig:frozen-defect}(a) shows $n_{ex,S}(b\rar 0)$ vs. $\gamma/w$  by Eq.~\eqref{eq:nex} for four different $S$ from $S=1/2$ to $S=2$. As in the $k$-resolved result, a curve with a larger $S$ is always above another curve with a smaller $S$. For each value of $S$, $n_{ex,S}(b\rar0)$ increases monotonically with $\gamma$, and it experiences a singularity 
at $\gamma=w$ where the slope of the curve (namely, $\partial n_{ex,S}(b\rar 0)/\partial \gamma$) changes discontinuously. (Due to the arcsin term in Eq.~\eqref{eq:nex}, this slope diverges as $\gamma\rar w^-$; it is finite as $\gamma\rar w^+$.) The physical reason of this singularity is the same as in the $S=1/2$ case, namely, $\gamma=w$ this is the critical point between the two cases that a portion of or all of the $k$ sectors traverse the $\mathcal{PT}$-broken region. Using $P_{n}^{(\alpha ,\beta )}(-1)=(-1)^{n}{n+\beta \choose n}$, one obtains the excitation density at $\gamma=w$ as $n_{ex,S}(b\rar0)
=1-  {2S-1/2  \choose \lfloor S \rfloor}/2^{2S}$. 
At $\gamma>w$, unlike the $S=1/2$ case where $n_{ex,S}(b\rar0)$ stays constant, for larger $S$, $n_{ex,S}(b\rar0)$ keeps increasing. For very large $\gamma$, from \eqref{eq:nex} we see that $n_{ex,S}(b\rar0)$ saturates at the value $1 - 1/2^{N - 1}=1 - 1/2^{2S}$.  As $S$ becomes larger, this saturation value approaches $1$. This means that for large $S$, at large non-Hermiticity the probability to stay in the ground state goes to zero in the adiabatic limit, indicating strong defect freezing.

The expressions of $n_{ex,S}(b\rar0)$ in terms of special functions may still seem opaque. Below we consider the case when non-Hermiticity is small, namely, when $\gamma\ll w$, and show that $n_{ex,S}(b\rar0)$ can be approximated by a very simple form. Since $F_1(a,b_{1},b_{2};c;x,y)\rar1 $ at $x\rar0$ and $y\rar 0$, we have
\begin{align}\label{eq:nex-linear}
&n_{ex,S}(b\rar 0,\gamma\ll w)\approx\frac{2}{\pi}\left[\arcsin\frac{\gamma}{w}-\frac{1}{(2S+1)\sqrt{\frac{w^2}{\gamma^2}-1}}\right]\nn\\
&\approx\frac{2}{\pi}\left(1-\frac{1}{2S+1}\right)\frac{\gamma}{w}.
\end{align}
Therefore, at small non-Hermiticity, $n_{ex,S}(b\rar 0)$ increases linearly with non-Hermiticity, and the proportionality constant depends on $S$ as $1-1/(2S+1)$, which increase from $1/2$ at $S=1/2$ to almost $1$ at large $S$. In Fig.~\ref{fig:frozen-defect}(b) these linear approximations are shown as dashed lines. They agree well with the exact results (the solid lines) at small $\gamma$; visible deviation appears roughly above $\gamma=0.2$.


\subsubsection{Scaling above the frozen defects}

Up to now we have been focusing on the adiabatic limit $b\rar 0$. We now consider small but finite $b$. We evaluate the quantity $\Delta n_{ex,S}=n_{ex,S}-n_{ex,S}(b\rar0)$, namely, the excitation density above the frozen defects in the adiabatic limit. We expect that $\Delta n_{ex,S}$ goes to zero as $b\rar 0$, and it should obey a certain scaling at finite $b$. 
Below we fix to the $\gamma<w$ region. We have
\begin{align}\label{}
&\Delta n_{ex,S}
=\frac{1}{\pi} \int_{-\frac{\pi}{2}}^{\frac{\pi}{2}}  [p_{k,S}-p_{k,S}(b\rar 0)] dk.
\end{align}
Numerics show that major contribution to the integral lies in a small region to the right of $k=\arcsin(\gamma/w)$
, and $p_{k,S}-p_{k,S}(b\rar 0)=1$ slightly above $k=\arcsin(\gamma/w)$. It is convenient to make a change of variable 
$x=\sqrt{w^2-\gamma^2}[k-\arcsin(\gamma/w)]$. Then near $k=\arcsin(\gamma/w)$ one has $\gamma+w\sin k\approx 2\gamma$, $\gamma-w\sin k\approx -x$, and $\delta\approx \gamma x/b$. So we can approximate the integral by
\begin{align}\label{}
& \int_{-\frac{\pi}{2}}^{\frac{\pi}{2}}  [p_{k,S}-p_{k,S}(b\rar 0)] dk\nn\\
&\approx  \int_{\arcsin\frac{\gamma}{w}}^{\frac{\pi}{2}}\left\{1- \left[1-\frac{ \gamma+w\sin k}{2\gamma-(\gamma-w\sin k) e^{2 \pi\delta}}\right]^{2S}\right\} dk\nn\\
&\approx \frac{1}{\sqrt{w^2-\gamma^2}}\int_{0}^{\infty}\left[1- \left(1-\frac{ 2\gamma}{2\gamma+xe^{\frac{2 \pi  \gamma x}{b}}}\right)^{2S}\right]dx.
\end{align}
This integral seems difficult to be performed analytically, but it allows a good approximation by simply finding the width at half maximum of the integrand. 
This width corresponds to an $x$ satisfying the equation
\begin{align}\label{}
1- \left(1-\frac{ 2\gamma}{2\gamma+x e^{\frac{2 \pi  \gamma x}{b}}}\right)^{2S}=\frac{1}{2},
\end{align}
or
\begin{align}\label{}
\frac{x e^{\frac{2 \pi  \gamma x}{b}}}{2\gamma} = \frac{1}{2^{\frac{1}{2S}}-1},
\end{align}
whose solution is
\begin{align}\label{}
&x=  \frac{b W_0\left( \frac{4 \pi  \gamma^2 }{(2^{\frac{1}{2S}}-1)b}\right)}{2 \pi  \gamma  },
\end{align}
where $W_0(x)$ is the Lambert $W$ function (the product logarithm) defined as the solution of $ ye^{y}=x$ at $y>0$. Using this value of $x$ to approximate the result of the integral, we get an analytical approximate expression on the excitation density above the frozen defects:
\begin{align}\label{eq:nex-W}
&\Delta n_{ex,S} \approx\frac{b W_0\left( \frac{4 \pi  \gamma^2 }{(2^{\frac{1}{2S}}-1)b}\right)}{2 \pi^2 \gamma \sqrt{w^2-\gamma^2}   }.
\end{align}
$W_0(x)$'s asymptotic expansion at large $x$ reads $W_{0}(x)=\ln x-\ln \ln x+(\ln \ln x)/\ln x+\ldots$ Therefore, for a general $S$, $\Delta n_{ex,S}$ scales with $b$ approximately as
\begin{align}
\Delta n_{ex,S}\propto b  \ln \frac{1}{ b} = b|\ln b| 
\end{align}
at small $b$, namely, a linear dependence with a slowly-varying logarithmic factor. 
In Fig.~\ref{fig:frozen-defect}(c), we plot $\Delta n_{ex,S}$ vs. $b$ from exact numerical integration and by the approximate expression \eqref{eq:nex-W} for two values of $S$ ($S=1/2$ and $S=2$) at $w=1$ and $\gamma=0.3$ (we also performed calculations at other values of $S$ and $\gamma$, and the results are similar), and find good agreement at small $b$. 
At larger $b$, the approximation \eqref{eq:nex-W} drops below the exact result; this is reasonable since in obtaining \eqref{eq:nex-W} we approximated the integral by the contribution from the region to the right of $k=\arcsin(\gamma/w)$, whereas as $b$ becomes larger the contribution from the region to the left of $k=\arcsin(\gamma/w)$ becomes more significant.

\subsection{Discussions}

We first summarize the main results in this section for defect freezing in the spin-$S$ $\mathcal{PT}$-SSH model \eqref{eq:n-SSH-S} under a quench $v=bt$.
For a two-band model like the $\mathcal{PT}$-SSH model, previous studies \cite{Sim-2023,Deng-2025} have shown that defect freezing in a $k$ sector takes place only when it is quenched through the $\mathcal{PT}$-broken region, or equivalently, through a pair of EPs. Here our results suggest that for the spin-$S$ $\mathcal{PT}$-SSH model (a multi-band model), the condition of existence of defect freezing in a $k$ sector is the same, namely, it needs to traverse the $\mathcal{PT}$-broken region (or pass through a pair of higher-order EPs) during the quench. The excitation density in the adiabatic limit increases with $S$, indicating that defect freezing is more significant for larger $S$. Our analytical results also reveal several quantitative properties of defect freezing. Namely, at small non-Hermiticity $\gamma$, the excitation density increase linearly with $\gamma$. At $\gamma=w$, the excitation density experiences a singularity, which is due to the fact that $\gamma=w$ is the critical point that separates two cases in which only a part of or all of the $k$ sectors pass the $\mathcal{PT}$-broken region. At small but finite quench rate $b$, the excitation density above the frozen defects scales with the quench rate $b$ approximately as $b\ln(1/ b)$.

Second, we recall that the analytical results of defect freezing across higher-order EPs in this section is obtained by applying the general result \eqref{eq:Us-result} or \eqref{eq:Us-result-2} in Sec. II which connects a spin-$S$ model to a spin-$1/2$ model. If such analytical results were absent, of course the problem can still be attached by direct numerical simulations of Schr\"{o}dinger equations, but then some properties of defect freezing may become opaque, for example, the singularity at $\gamma=w$ and the scaling above the frozen defects. This shows the usefulness of the results in Sec. II, especially when taking into account the fact that most time-dependent non-Hermitian problems 
do not allow such an exact analytical treatment.


Finally, we discuss possible experimental realizations to observe defect freezing across higher-order EPs considered in this section. The $\mathcal{PT}$-SSH model, as sketched in Fig.~\ref{fig:nSSH}(a), has been realized on different physical platforms, for example, photonic lattices \cite{Weimann-2017,Qian-2024} 
and electric circuit networks \cite{Stegmaier-2021}. 
Since the spin-$S$ $\mathcal{PT}$-SSH models simply differ from the $\mathcal{PT}$-SSH model by adding more lattice sites (as examples, see the $S=1$ and $S=3/2$ cases sketched in Fig.~\ref{fig:nSSH}(b) and (c)), we expect that they are also readily realizable on these platforms. To perform time-dependent drives of hopping amplitudes between lattice sites, the electric circuit platform is perhaps more convenient. In an electric circuit network \cite{Stegmaier-2021}, hoppings between sites are realized by capacitors, and on-site gain and loss are realized by resistive elements. Time-dependent manipulation of hoppings can then be achieved by using variable capacitors, for example those based on varactors \cite{Taravati-2017}. In a photonic lattice system made of waveguides \cite{Weimann-2017}, hoppings between sites are controlled by distances between waveguides, and on-site imaginary potentials can be engineered by wiggling the waveguides to introduce radiative loss. Time-dependent manipulation of hoppings can be introduced by motions of waveguides, which may require sophisticated mechanical control. 


\section{Conclusions and outlooks}

Using the Wei-Norman approach, we show that for a spin-$S$ under a general non-Hermitian time-dependent drive, the elements of its evolution operator can be expressed in terms of those of the corresponding spin-$1/2$ model in closed forms via Jacobi polynomials (Eqs.~\eqref{eq:Us-result} or \eqref{eq:Us-result-2}), which means that from any solvable two-level non-Hermitian quantum model one can construct a class of solvable multi-level models. This approach is straightforward and friendly to readers with little knowledge on Lie algebra. We further apply this result to investigate the phenomenon of defect freezing (i.e. the breakdown of adiabaticity) in nonequilibrium non-Hermitian systems. We construct the so-called spin-$S$ $\mathcal{PT}$-SSH models, which are spin-$S$ extensions of a $\mathcal{PT}$-symmetric SSH model, and obtain analytical expressions for excitation probabilities under linear quenches. We find that for these models defect freezing occurs in the momentum sectors that pass through a pair of higher-order EPs during the quench. Several quantitative results are also revealed, namely, the excitation density increases linearly with non-Hermiticity $\gamma$ at small $\gamma$ and experiences a singularity at a specific $\gamma$; at a finite quench rate $b$, the excitation density above the frozen defects approximately scales as $b\ln(1/ b)$. Our work 
presents a general method to construct analytically solvable multi-level nonequilibrium non-Hermitian models, and predicts analytical results of defect freezing across higher-order EPs which can possibly be tested experimentally in electric circuit networks or photonic lattices.

The current study might be extended in the future as follows. For the spin-$S$ $\mathcal{PT}$-SSH models under quenches, since our interest is in defect freezing, we focussed on the adiabatic limit $b\rar 0$ or small $b$ (namely, when the quenches are slow). But since the analytical expression of excitation probabilities \eqref{eq:pks} at arbitrary parameters (with $p_{k,1/2}$ given by \eqref{eq:pk-1o2}) is available, one may also analyze excitations in more general cases at any $b$, e.g., the behavior of transition probabilities at large $b$ (namely, for fast quenches). Besides, we fixed to the situation that the initial state in each $k$ sector is its ground state. One may consider more general initial states, for example the ``central'' state for an odd $S$ with $m=0$, though the expressions of elements of evolution operators would be more complicated than \eqref{eq:U-S-j1} for the case of ground states as initial states. Another future direction is to incorporate the metric formalism \cite{Sim-2023} into the study of defect freezing across higher-order EPs. In our work, we employed the direct normalization for the probabilities, which is commonly used in studies of non-Hermitian systems. Recently, Sim et al. \cite{Sim-2025} proposed that, compared to a direct normalization, the metric formalism is more suitable for non-Hermitian systems where non-Hermiticity can be directly engineered. Since the non-Hermitian LZ model is still exactly solvable in the metric framework \cite{Sim-2023}, we expect its spin-$S$ extension to be solvable as well in this framework, and it would be interesting to derive this solution and analyze defect freezing across higher-order EPs in the metric formalism. Finally, one may also try to apply the general result of reduction of a non-Hermitian spin-$S$ model to a spin-$1/2$ model, i.e., Eqs.~\eqref{eq:Us-result} or \eqref{eq:Us-result-2}, to a broader range of nonequilibrium non-Hermitian systems. In principle, whenever a two-level non-Hermitian system is exactly solvable (e.g. those studied in \cite{Torosov-2017,Shen-2019,Longstaff-2019,Sim-2023,Malla-2023,Wang-2022,Pan-2024,Hu-2025,Luo-2017,Liu-2024}), so is its corresponding spin-$S$ model. 
For example, one may investigate spin-$S$ extensions of the non-Hermitian Rosen-Zener model or non-Hermitian periodically driven (Floquet) systems which are analytically solved in \cite{Luo-2017}, and hopefully gain new physical understanding from the analytical results of these multi-level models.

\section*{Acknowledgements}
We thank Fuxiang Li for helpful discussions. This work was supported by the National Natural Science Foundation of China under Grant No. 12105094, and by the Fundamental Research Funds for the Central Universities from China.


\begin{thebibliography}{}
\bibitem{note-stationary} In contrast, for a time-independent (satationary) quantum Hamiltonian, finding its exact dynamics amounts to analytically solving its eigenvalues and eigenstates, a task that is readily achievable at least for two-level systems.
\end{thebibliography}
\end{document}